# OACAL: Finding Module-consistent Specifications to Secure Systems from Weakened User Obligations

Pengcheng Jiang
*Waseda University*
Tokyo, Japan
rexjiang@fuji.waseda.jp

Kenji Tei
*Waseda University/*
*National Institute of Informatics*
Tokyo, Japan
ktei@aoni.waseda.jp

*Abstract*—Users interacting with a system through UI are typically obliged to perform their actions in a pre-determined order, to successfully achieve certain functional goals. However, such obligations are often not followed strictly by users, which may lead to the violation to security properties, especially in security-critical systems. To improve the security with the awareness of unexpected user behaviors, a system can be redesigned to a more robust one by changing the order of actions in its specification. Meanwhile, we anticipate that the functionalities would remain consistent following the modifications. In this paper, we propose an efficient algorithm to automatically produce specification revisions tackling the attack scenarios caused by weakened user obligations. By our algorithm, all the revisions would be generated to maintain the integrity of the functionalities using a novel recomposition approach. Then, the eligible revisions that can satisfy the security requirements would be efficiently spotted by a hybrid approach combining model checking and machine learning techniques. We evaluate our algorithm by comparing its performance with a state-of-the-art approach regarding their coverage and searching speed of the desirable revisions.

## I. INTRODUCTION

Nowadays, with the rapid development of software applications for various uses on a ubiquitous range, people can hardly live without the convenience they provide. According to recent studies [1, 2], people today prefer booking flights and hotel rooms through online applications rather than making phone calls, because the software provides more visual information that can be accessible as long as the Internet is connected. They may also stay in a particular state for a long time in order to ponder and make more cautious decisions. Thus, systems that place less constraints on user behavior while maintaining security properties are in demand. However, most software systems do not take additional considerations regarding user obligation and how its weakening could threat the security on the user's end. User obligation refers to a set of expectations for a user's actions when using software, which are commonly defined in specifications that can be modeled by transitional systems [3]. Nevertheless, users are not programs that can always follow instructions to the letter. A user may deviate from the expected behaviors and temporally leave the machine in locations where others can gain access to it, which can be either offline (e.g., physical access) or online (e.g., Remote Access Trojan) [4]. Such deviation leads to the violation of security requirements of the system, since that malicious users who cannot be identified may take control of the machine and operate on the applications unauthorizedly, which is not expected. To secure the system from this potential threat, either the user behavior should be regulated, or the system's specification should be re-designed. In this study, we concentrate on the latter methodology since we reckon that it is the developer's responsibility to make the security system more resilient and user-friendly.

OASIS is a state-of-the-art approach proposed by Tun et al. (2020) [5] to automatically generate revisions of the specification by an iterative abstraction and synthesis algorithm. In our integrated study, we highlight four main limitations of the OASIS approach, which are careless of functional consistency, limited revision coverage, low searching speed and lack of requirement degradation.

In this paper, we propose an efficient algorithm – OACAL (Obligation, Attack scenario, model Checking And machine Learning) to automatically generate specification revisions that maintain the consistency of functionalities in their sub interfaces using two major approaches: module-based recomposition and accelerated checking, which are distinct from abstraction-based recomposition and controller synthesis in OASIS respectively. Besides, requirement degradation is added as an optional strategy for specification selection. For implementation, we use an existing model checker *LTSA* [6] for model checking and *Turi Create* [7] as a framework for the machine learning process. The evaluation is conducted as a comparative study between OASIS and OACAL in terms of the impact of their different revision coverage and the searching speed for the eligible specifications (i.e., the revisions that can satisfy the requirements). In addition, we theoretically demonstrate the effectiveness of our approach.

The remainder of this paper is organized as follows. Section II gives definitions and explanations of some terminologies as the background. Section III firstly introduces several related works, then focuses on the OASIS approach and reveals some of its shortcomings based on our study. Section IV explains our proposal in detail. Section V evaluates our approach by conducting a case study and an efficiency analysis, before stepping up to the discussion of the findings with theoretical proof. At the end, Section VI concludes by summarizing the paper and outlining potential future works.

## II. BACKGROUND

In this section, some concepts are introduced as the preliminaries to comprehensively understand our work. In this paper, Labeled Transition Systems (LTS) [8] are used to formalize the system specifications and environments.

*Definition 1.* (Labelled Transition System) An LTS $\mathcal{M}$ is a 4-tuple $\mathcal{M} =< S, s_0, \mathcal{A}, \Lambda >$, where:
- $S$ is a set of states
- $s_0 \in S$ is the initial state
- $\mathcal{A}$ is a set of actions
- $\Lambda \subseteq S \times \mathcal{A} \times S$ is a set of transition relations

**Example.** We give an example of a ticket-booking system expressed in LTS, as presented in Fig. 1.

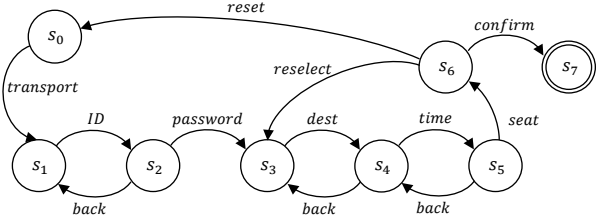

Fig. 1. An LTS example: ticket-booking system.

In this example, state set $S = \{s_0, s_1, ..., s_7\}$ containing the initial state $s_0$, action set $\mathcal{A} = \{transport, ID, password, ..., confirm\}$, transition relation set $\Lambda = \{s_0 \xrightarrow{transport} s_1, s_1 \xrightarrow{ID} s_2, ..., s_5 \xrightarrow{confirm} s_6\}$. Actions are the abbreviations of "select transport (train, steamer or flight)", "input personal identification", "input password", "select destination", "select departure time", "select seat", "confirm the reservation", "reselect destination", "reset state" and "go back".

To check the security and the functionalities of the specifications, we formalize both safety and functional requirements by Linear Temporal Logic (LTL) [9]. LTL is composed by a finite set of atomic propositions, temporal modal operators and logical operators [10, 11]. Thus, an LTL $\phi$ is defined as $\phi ::= \neg \phi \mid \phi \vee \varphi \mid \phi \wedge \varphi \mid \phi \rightarrow \varphi \mid \phi \leftrightarrow \varphi \mid true \mid false \mid X\phi \mid \phi U \varphi \mid \phi W \varphi \mid \phi R \varphi \mid \phi M \varphi \mid F\phi \mid G\phi$ where $\varphi$ is another LTL; $\neg$ (negations), $\vee$ (or), $\wedge$ (and), $\rightarrow$ (implication), $\leftrightarrow$ (equivalence), true and false are logical operators; $X$ (next, $\bigcirc$), $U$ (until), $W$ (weak until), $R$ (release), $M$ (strong release), $F$ (finally) and $G$ (globally, $\square$) are temporal modal operators. In LTL, we describe the satisfaction relation with the symbol $\vDash$. In the LTL model checking with an LTS model $\mathcal{M}$, we say $\mathcal{M} \vDash \phi$ if all paths in the $\mathcal{M}$ do not violate LTL property $\phi$ [12]. The process of LTS-LTL model checking through LTSA could be visualized by Fig. 2.

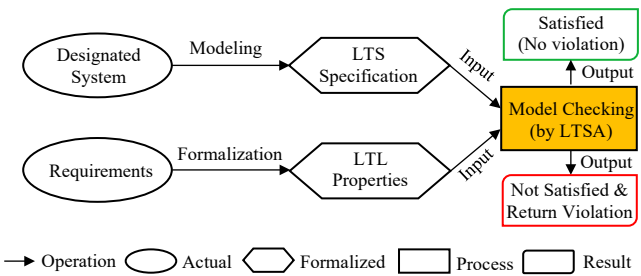

Fig. 2. The model checking process through LTSA.

To express the consistency of the functionalities in each interface of a specification, we define an integration of them called module.

*Definition 2.* (Module) A module $m$ is a set of all the actions at a specific state in LTS, and $m \subseteq \mathcal{A}$, which could be used to claim all serviceability on an interactable interface.

**Example.** We take the example in Fig. 1 to illustrate this definition. In Fig. 1, we have seven modules in total, which are: $m_0 = \{transport\}$, $m_1 = \{ID\}$, $m_2 = \{password, back\}$, $m_3 = \{dest\}$, $m_4 = \{time, back\}$, $m_5 = \{seat, back\}$, $m_6 = \{confirm, reselect, reset\}$.

Module plays a significant role in proceeding our work, and its involvement into the modeling of software specification should also be defined.

*Definition 3.* (Modularized Labelled Transition System) Regarding module, we add two attributes to the LTS to make it a tuple $\widetilde{\mathcal{M}} =< \mathcal{M}, \widehat{m}, d >$, called as *mLTS*, where:
- $\mathcal{M}$ is a set of LTS
- $\widehat{m}$ is a set of modules
- $d = \{+, -, \xrightarrow{s}, \xrightarrow{m}\}$ is a set of directions of $\mathcal{A}$

Two mLTSs $\widetilde{\mathcal{M}}$ and $\widetilde{\mathcal{N}}$ are claimed to be **module-consistent** if $\widehat{m}_{\widetilde{\mathcal{M}}} = \widehat{m}_{\widetilde{\mathcal{N}}}$, regardless of the sequences of their modules. The action classes of $\widehat{m}$ are the same as $\mathcal{A}$, as any actions should be performed in a module. The directions for all actions must be pre-declared in order to analyze module consistency; otherwise, the consistency would be ambiguous. In an mLTS, we declare the type of an action based on its direction $d$:

a) Forward-stepwise action $\mathcal{A}_+$: an action set that transitions the current state $s_t$ to the next state $s_{t+1}$. For the ticket-booking example in Fig. 1, $\mathcal{A}_+ = \{transport, ID, password, dest, time, seat, confirm\}$.

b) Backward-stepwise action $\mathcal{A}_-$: an action set that transitions the current state $s_t$ to the previous state $s_{t-1}$ for $t \geq 1$, like action $\{back\}$ in the example.

c) State-targeted action $\mathcal{A}_{\xrightarrow{s}}$: an action set that transitions the current state $s_t$ to a fixed target state $s_{t'}$. In the example, action $\{reset\}$ transitions the state to $s_0$.

d) Module-targeted action $\mathcal{A}_{\xrightarrow{m}}$: an action set that transitions the current state $s_t$ to the state $s_m$ of a target module $m$. In Fig. 1, action $\{reselect\}$ transitions the state to the state of a target module that has the action $\{dest\}$.

## III. RELATED WORKS AND PROBLEMS

In this section, we provide a literature review and point out some existing challenges, to explicitly manifest the motivation and contribution of our work.

*Security and usability.* It is no doubt that both security and usability are essential to the development of software so that neglecting either of them renders the software ineffectual. However, a perceived trade-off between security assurance and ease of use makes the task of merging security and usability challenging to developers [13]. As most conventional security-critical systems have shown their apathy about user experience (e.g., setting complex password) and overburdened users, users may lose incentive to comply with the systems, leaving security policies ineffective [14]. Studies tackling this issue involve the simplification of authentication [15-17]; the prediction of human-computer interaction (HCI)

[18] and using HCI security principles [19] for user-interface designing; the parallel analysis of elicited security concerns and usability confliction [20]; and so on. Nonetheless, based on our investigation, none of those approaches have further discussion on the weak assumption of the environment where users interact with the system, although in early years Sasse et al. (2001) [21] and Bishop (2005) [22] had pointed out that the security configuration could become invalid by changing physical environments.

*Specification revision.* When considering the external environment of the system, we need to model it in the same manner as the machine behavior so that they can be composed together, based on which we could find a new specification as the solution. The existing approaches to find a revision $M_r$ of a specification $M_o$ that can meet the given requirements could be generalized to three types [5], which are:

1) Direct controller synthesis that generates a specification as a refinement of the original one ($M_o$ refines $M_r$) with a subset alphabet ($\mathcal{A}_{M_r} \subseteq \mathcal{A}_{M_o}$) [23, 24].
2) Add or remove actions to the original specification by changing the environment model (e.g., take additional authentication [25, 26]) so that $M_o$ does not refine $M_r$ and $\mathcal{A}_{M_r} \nsubseteq \mathcal{A}_{M_o}$.
3) Recompose the specification (make different sequence of actions) to a new one with the original alphabet (e.g., the OASIS approach) so that $M_o$ does not refine $M_r$ and $\mathcal{A}_{M_r} \subseteq \mathcal{A}_{M_o}$.

Same as the OASIS approach, our focus is type 3 mentioned above. We will explain this sophisticated approach and specify some limitations of it.

*The OASIS approach.* The OASIS approach is proposed to automatically revise the specification to a more robust one that maintains the security of the system even when a user's behavior is deviated from the obligation. It is based on an iterative abstraction and synthesis method, which is depicted in Fig. 3. The algorithm starts from an original specification $M_o$ and ends with a revision $M_r$ or *Fail*. The details of OASIS are explored as follows.

*Step 1.* Select the next minimal subset $L$ from an ordered power set $\mathcal{L} = (2^{\mathcal{A}_m}, \subseteq)$ where $\mathcal{A}_m$ denotes action numbers in the alphabet of $M_o$ so that the smallest element is $\emptyset$ and the largest is $M_o$.

*Step 2.* Carry out hiding operations ("\" and "@") [27] with $L$ to obtain two abstractions $P$ and $Q$ of $M_o$. Then, compose two abstractions to $F$ by parallel composition ("||") [28].

*Step 3.* Compose $F$ and $E_2$ and put together with requirement model $R_s$ into controller synthesis where $E_2$ is the new environment model obtained by the parallel composition of the old environment $E_1$ model and the attack scenario model $W$.

*Step 4.* Check if a controller $C$ can be synthesized or not. If yes, it outputs the controller and returns the revision $M_r$ by the parallel composition of $C$ and $P$. If no, it further checks if the current $L$ is the last subset ($M_o$) or not. If no, it turns back to *Step 1*. If yes, it returns *Fail* as the result.

Based on our study, we discovered some problems of this approach, which are discussed as follows.

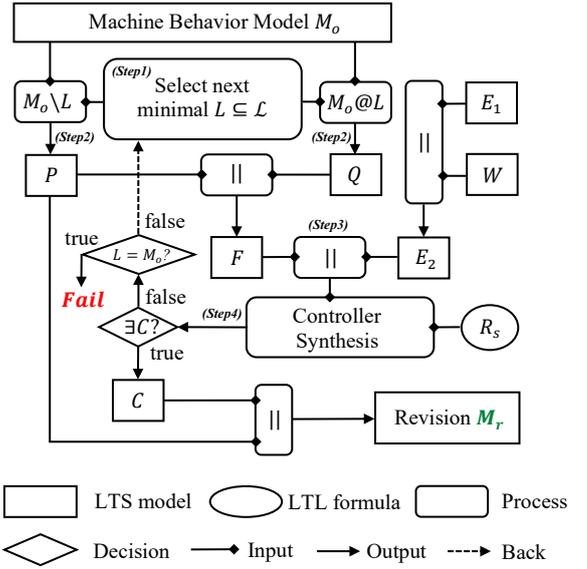

Fig. 3. The architecture of the OASIS approach

**Problem 1**: Careless of functional consistency. The principle of controller synthesis allows OASIS to remove some actions from the original specification, which may cause the loss of essential functionalities in complex systems.

**Problem 2**: Low coverage of revisions. Since the recomposition method in OASIS is abstraction-based, all of the revisions it generates are composed by two refinements of the original specification, which covers limitedly compared to the complete permutation of actions.

**Problem 3**: High computational time. Due to its iterative feature, OASIS has an exponential complexity $\mathcal{O}(2^m)$ where $m$ is the size of the specification alphabet. Moreover, there are many redundant revisions during the iterations since that the different subsets $L$ would generate plenty of common paths in $F$ that interleaved by $P$ and $Q$. We examined this redundancy regarding the module-consistent revisions covered by OASIS, which is shown in Table I. In the table, different rows show the cases of specifications with varying numbers of modules, while different columns present how redundant revisions are distributed across the selected subset $L$ from smallest to largest. Because of the symmetrical nature of the abstraction approach, we only need to check half of all subsets. For example, when the number of modules is 4, the result in column 3 is the same as the result in column 1.

TABLE I.     Distribution of Non-redundant Revisions by OASIS

| Modules | Element numbers of the selected *L* | | | | |
|---|---|---|---|---|---|
| | *1* | *2* | *3* | *4* | *Sum* |
| 4 | 10[a]/16[b] | 4/18 | -[c] | - | 14/34 |
| 5 | 17/25 | 25/100 | - | - | 42/125 |
| 6 | 26/36 | 79/225 | 27/200 | - | 132/461 |
| 7 | 37/49 | 188/441 | 204/1225 | - | 429/1715 |
| 8 | 50/64 | 380/784 | 766/3136 | 234/2450 | 1430/6434 |
| 9 | 65/81 | 689/1296 | 2158/7056 | 1950/15876 | 4862/24309 |

[a.] The numerator denotes the number of the revisions having no appearance before.
[b.] The denominator denotes the number of all revisions generated with *L*.
[c.] The dash denotes the duplicated part that is unnecessary to be examined.

In Fig. 4, we also present the indices of the median and the last non-redundant revisions, which can be interpreted as the theoretical searching time in the average case and worst case, respectively. By our observation, such redundancy has more negative impact on finding eligible revisions when the original specification contains more modules.

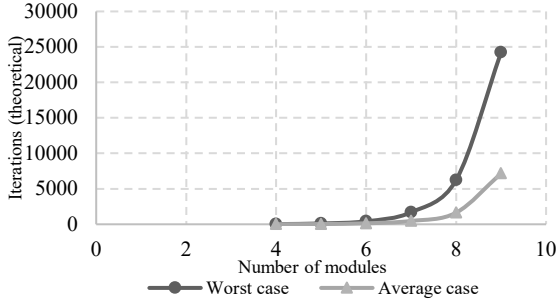

Fig. 4. Theoretical iterations needed to find eligible revisions by OASIS

**Problem 4**: No requirement degradation. OASIS generates specifications based on whether or not a controller can be synthesized. However, when there are several security requirements weighted differently by priority and those that are weighted relatively lowly are optional to be satisfied, OASIS is no longer effective since that it only returns the specification that can perfectly satisfy all requirements. Therefore, the OASIS approach might be inefficient for developers when they have numerous requirements for the system but have no idea that which of them could probably be satisfied together. In that circumstance, developers require requirement degradation, which allows them to "hope for the best" [29] and at least acquire some suboptimal revisions.

## IV. THE OACAL APPROACH

In this section, we give a stepwise introduction of our algorithm – OACAL giving solutions to the aforementioned problems of OASIS. The architecture of our approach is presented in Fig. 5. In OACAL, module-based recomposition ensures that all functionalities in the original specification are preserved after the revision, and provides a complete coverage of such revisions. The accelerated checking approach then efficiently finds the eligible ones among those revisions. Finally, for specification selection, requirement degradation could be applied if the spotted revisions are undesirable. More details of our approach are explained as follows.

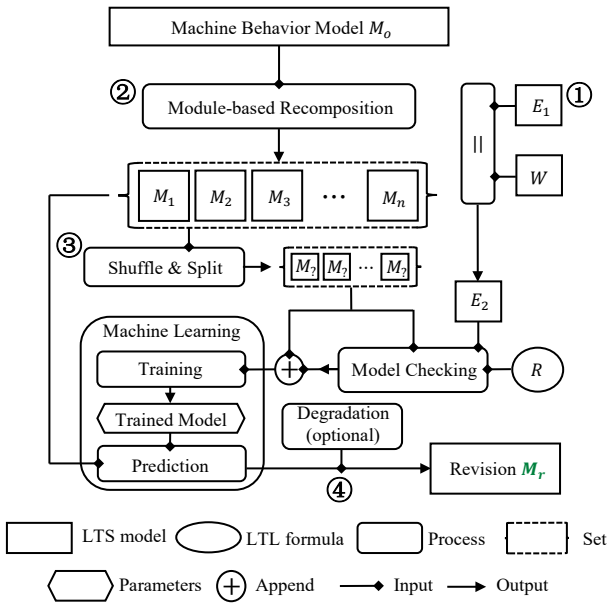

Fig. 5. The architecture of the OACAL approach

①*Weakening with attack scenario*. Similar to the OASIS approach, the input of OACAL also contains three models: (1) a machine behavior model $M_o$, (2) an environment model $E_2$ and (3) a requirement model $R$. The only difference is that the $R$ here is composed of both security requirement $R_s$ and functional requirement $R_f$, whereas OASIS only has input $R_s$ since that it maintains the functionalities of the specification by its abstraction approach. As the specification and requirements are the given conditions, we need to deliberate the attack scenario $W$ that may occur due to the weakened user obligations so that $E_2 \leftarrow E_1 || W$ and $E_2 || M_o \not\models R$ where $E_1$ is the original environment (usually same as $M_o$). For this step, we take the "weakened" environment $E_2$ as the output.
**Example.** Consider a system in an unsafe environment on the left of Fig. 6, some external behaviors such as "enter room" and "exit room" could be regarded as weakening condition to the environment, as we have discussed. For instance, if we compose the LTS model of these behaviors with the ticket-booking system specification in Fig. 1, some security requirements such as "no privacy leakage" would be violated since that the action sequences like "$u.ID \rightarrow u.exit \rightarrow mu.enter \rightarrow mu.back$" exist and make threats.

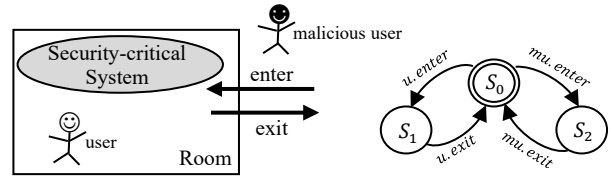

Fig. 6. An example of weakening condition (left: actual scenario, right: LTS model, $u$: user's behavior, $mu$: malicious user's behavior).

②*Module-based recomposition*. A unique recomposition approach plays an essential role in the algorithm of OACAL. It is called module-based recomposition since that the unit being used in partitioning the original LTS specification is no longer the action, but the module. To remove redundancy and maintain functional consistency, we only change the sequence of the modules in the original specification but do not make any modifications inside any of them. The module consistency thus would be guaranteed by this approach. Formally speaking, given an original behavior model by an mLTS $M_o$ with a module set $\widehat{m}_0$ and the recomposed specifications $M_1$, $M_2$, …, $M_n$ with the module sets $\widehat{m}_1$, $\widehat{m}_2, …, \widehat{m}_n$, we have $\widehat{m}_0 = \widehat{m}_1 = \widehat{m}_2 = \cdots = \widehat{m}_n$. Therefore, the number of revisions $n$ would be the complete permutation of the modules so that $n = \text{num}(\widehat{m}_0)!$ For example, when $M_o$ has five modules, there would be 5! = 120 non-redundant module-consistent revisions generated by module-based recomposition, whereas OASIS only provides 42 revisions, as revealed in Table I. The output of this step is a set of revisions ranging from $M_1$ to $M_n$.
**Example**. Fig. 7 describes how we decompose the specification of the ticket-booking system (in Fig. 1) into modules and recompose a new specification. There would be a total of 7! = 5040 revisions in this scenario because there are seven modules. Here, the 5040[th] revision is used as an example.

③*Accelerated checking*. In OACAL, we combine model checking and machine learning to make the checking process more efficient. This paper is not the pioneering work combining these two techniques, and most of the existing works have proved the feasibility of this approach with their remarkable results [30-33]. Zhu et al. [33] who applied machine learning to the LTL model checking with the specification modeled by Kripke Structures (KS), claimed that model checking is a binary classification task for machine learning. The 100% accuracy they achieved with Logistic Regression indicates that the process of model checking is highly learnable, which gives an inspiration to our work. Based on our survey, we are the first explorer implementing this idea on the LTS-LTL model checking. The mechanism of our accelerated checking approach is explained as follows.

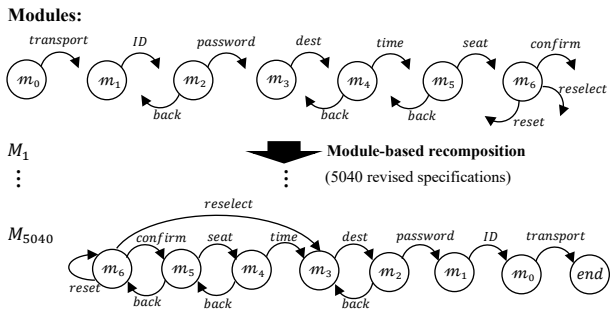

Fig. 7. Module-based recomposition on the ticket-booking system

**Step 1.** Shuffle the revision set that we obtained by module-based recomposition and split out a training set from it.

**Step 2.** Carry out the model checking process to the revisions in the training set iteratively, with the environment $E_2$ and requirements $R$. Then, append the model checking results as the labels to the corresponding revisions to compose the training data.

**Step 3.** Train a machine learning model with the training data and a binary classification algorithm.

**Step 4.** Use the trained model to predict the model checking results for all specifications in the shuffled revision set. Then, output the predictions.

④*Specification selection*. Finally, we conduct the specification selection on all revisions labelled with the prediction results. The default outputs are the eligible revisions that satisfy all the requirements. However, if those revisions are unideal to be the new specification $M_r$, we may degrade the requirements so that some of the lower-priority ones can be waived as long as the overall payoff is fulfilled.

**Example.** Fig. 8 displays one of the optimal revisions of the ticket-booking system generated by the OACAL approach, which ensures the security requirements. With this specification, privacy (ID, destination, departure time, seat) would not be leaked even if the users deviate from the obligations and temporarily exit the room during the operation.

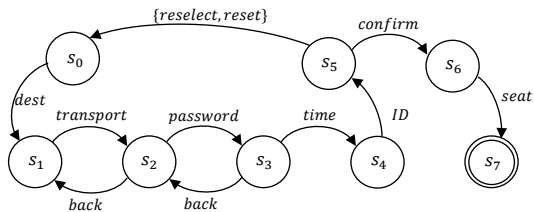

Fig. 8. A feasible revised specification for the ticket-booking system generated by OACAL (not in the coverage of OASIS)

## V. EVALUATION

In this section, we compare our proposal – OACAL with OASIS regarding their coverage and searching speed for module-consistent revisions. The following research questions are investigated.

**RQ1** How may OACAL generate more module-consistent revisions than OASIS with its larger coverage?
**RQ2** How can requirement degradation be effective in specification selection?
**RQ3** How fast is OACAL on finding revisions with its accelerated checking mechanism?

### A. Settings

*Platform.* The platform we use is described as follows.
1) CPU: Intel(R) Core (TM) i7-6700K @ 4.00GHz.
2) RAM: 8.0 GB.
3) OS: Windows 10 64-bit.
4) LTSA: LTS-LTL model checking tool.
5) JDK 1.6: Development environment for Java.
6) Jupyter Notebook: Development environment for Python.
7) Turi Create: Toolkit for machine learning algorithms.

*Experimental procedures.* For RQ1 and RQ2, we implement both OASIS and OACAL in a case study. Then, we compare the specification revisions generated for the example by both approaches. We also carefully examine how the requirement degradation mechanism works in that case. For RQ3, we set the revision coverage of OASIS as a common coverage for both OASIS and OACAL approaches. Then, we test the time consumption by each of them to search the first module-consistent revision that can satisfy the same LTL requirement randomly generated by a requirement generator, as presented in Fig. 9. The experiment is conducted six times for the specifications from 4 modules to 9 modules.

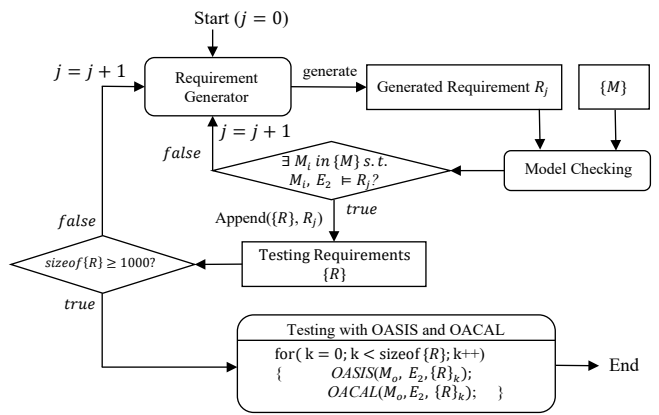

Fig. 9. Experimental setting for efficiency analysis of OASIS and OACAL ($\{M\}$: the set of the module-consistent revisions of $M_o$ in OASIS's coverage, $\{R\}$: a set of requirements that can be satisfied by at least one revision in $\{M\}$, the testing starts when the size of $\{R\}$ is 1000).

### B. Case study: Online Gifting System

Nowadays, with the rapid growth of the entertainment industry, online gifting system that enables users to send their virtual gifts to creators, becomes popular. This system is usually embedded in video-sharing or live-streaming software, and the virtual gifts are paid with real money. For instance, "Bits" in Twitch (a live-streaming website) [34], "Coins" in TikTok (a video-sharing and live-streaming mobile application) [35], "Niconico Point" in Niconico (a Japanese video-sharing website) [36], "B-Coin" in Bilibili (a Chinese video website) [37], and so on. All of them could be obtained immediately through a charge-and-pay process. Therefore, this system is considered security-critical since it directly relates to the safety of the user's personal asset. An example of a typical online gifting system is given in Fig. 10.

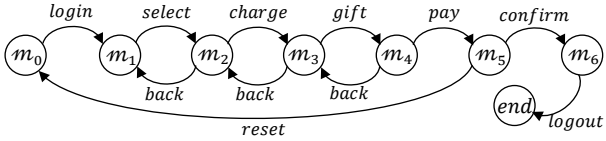

Fig. 10. The specification of an online gifting system example

In this example, the forward-stepwise actions $\mathcal{A}_+ = \{login, select, charge, gift, pay, confirm, logout\}$ are translated to the behaviors "log in", "select a receiver to gift", "charge to the account", "decide the number and category of the gift(s) to send", "make payment", "confirm" and "log out", respectively. There are also a backward-stepwise action $\mathcal{A}_- = \{back\}$ that returns to the previous state with the inputs remained and a state-targeted action $\mathcal{A}_{\underset{\rightarrow}{s}} = \{reset\}$ that removes all the inputs and back to the initial state. The weakening condition here is the same as the case in Fig. 6.

We place three security requirements and three functional requirements on this system, which are listed below with their weight ($w$) assigned. The word "a user" ($u$) refers to a normal user who is not "a malicious user" ($mu$).

*SR1: No change of the receiver (w=4).* Action $\{select\}$ should not be performed by a malicious user after it has been performed by a user. LTL: $\Box(u.select \rightarrow \Box(\neg mu.select))$.

*SR2: No leakage of the bill (w=4).* After the execution of action $\{charge\}$ by a user, action $\{back\}$ should not be taken by a malicious user if $\{charge\}$ is one of its next actions. LTL:$\Box(u.charge \rightarrow \Box(\neg mu.back \rightarrow \bigcirc mu.charge))$.

*SR3: No modification to the gift (w=3).* Action $\{gift\}$ should not be performed by a malicious user after its execution by a user. LTL: $\Box(u.gift \rightarrow \Box(\neg mu.gift))$.

*FR1: Charge after login (w=3).* Action $\{charge\}$ is placed after $\{login\}$. LTL: $\Box(charge \rightarrow \Box(\neg login))$.

*FR2: Pay after charge (w=2).* Action $\{pay\}$ is positioned after $\{charge\}$. LTL: $\Box(pay \rightarrow \Box(\neg charge))$.

*FR3: No confirmation after selection until gifting (w=1).* No execution of $\{confirm\}$ after $\{select\}$ until $\{gift\}$ is executed. LTL: $\Box(select \rightarrow (\neg\, confirm\; W\; gift))$.

The implementation details of OASIS and OACAL on this example are separately illustrated as follows.

*OASIS.* The OASIS approach could solely output one module-consistent revision satisfying all the requirements for this example, as presented in Fig. 11. However, it could not be considered as a desirable specification for the system. The revision places action $\{pay\}$ after action $\{logout\}$, which is impractical to implement in practice because most payments are done while the user's account is logged in. Furthermore, making a payment after logging out indicates that action $\{charge\}$ would be meaningless.

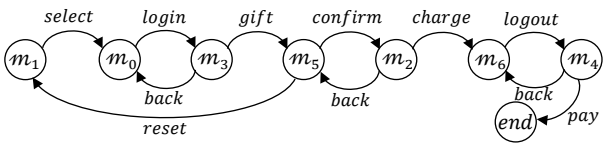

Fig. 11. The only module-consistent revision given by OASIS

*OACAL.* To implement the OACAL approach, a machine learning algorithm to handle binary classification should be specified. We used "*create*" function in Turi Create to run an automated evaluation of various typical algorithms for predicting checking results for all of the requirements, resulting in Table II. Note that the accuracy here are unideal since that this function only trains a few epochs for each algorithm in order to provide a quick comparison.

TABLE II. QUICK EVALUATION OF CLASSIFICATION ALGORITHMS BASED ON THE VALIDATION ACCURACY ON EACH REQUIREMENT

| Algorithm | Requirements ($R_s/R_f$: Security/Functional Requirement) | | | | | |
|---|---|---|---|---|---|---|
| | $R_{s1}$ | $R_{s2}$ | $R_{s3}$ | $R_{f1}$ | $R_{f2}$ | $R_{f3}$ |
| Boosted Trees | 95.24% | 89.68% | 93.25% | 97.22% | 98.41% | 89.68% |
| Random Forest | 94.44% | 84.52% | 80.56% | 92.86% | 94.84% | 82.54% |
| Decision Tree | 76.98% | 76.59% | 74.60% | 92.86% | 90.87% | 72.22% |
| SVM Classifier | 60.71% | 73.02% | 71.83% | 58.33% | 57.54% | 61.91% |
| Logistic Regression | 55.16% | 73.02% | 71.83% | 57.54% | 59.52% | 62.30% |

Gradient Boosted Trees (GBT) was clearly chosen as the model we utilized in OACAL due to its superior performance when compared to other models. We used 30% of the shuffled set's revisions as the training data and trained each requirement for 100 epochs. As a result, OACAL spotted 46 revisions out of 5040 module-consistent revisions that satisfied all of the requirements. We also confirmed that the prediction's overall precision and recall are both 1.0, showing the effectiveness of the module-based recomposition and the accelerated checking approach. Table III provides a more detailed evaluation using a variety of metrics.

TABLE III. PERFORMANCE OF GBT ON THE PREDICTION OF MODEL CHECKING RESULT

| Metrics | Trained Models for Requirements | | | | | |
|---|---|---|---|---|---|---|
| | $R_{s1}$ | $R_{s2}$ | $R_{s3}$ | $R_{f1}$ | $R_{f2}$ | $R_{f3}$ |
| Accuracy | 1.000 | 1.000 | 1.000 | 0.991 | 0.999 | 0.977 |
| AUC | 1.000 | 1.000 | 1.000 | 0.999 | 1.000 | 0.997 |
| Precision | 1.000 | 1.000 | 1.000 | 0.998 | 1.000 | 0.972 |
| Recall | 1.000 | 1.000 | 1.000 | 0.987 | 0.997 | 0.956 |
| F1-score | 1.000 | 1.000 | 1.000 | 0.992 | 0.999 | 0.964 |

From the generated 46 revisions, we filtered out two of them that have $\{logout\}$ as the final action, as shown in Fig. 12(a). Nevertheless, we found that the action $\{confirm\}$ becomes useless because it occurs before the action $\{select\}$, which is an alternative solution to fulfill FR3 but not our initial purpose of setting this requirement. Since FR3 has the lowest weight among all requirements, we could trade it off to check if better revisions exist. Fig. 12(b) shows two more desirable results obtained following the degradation.

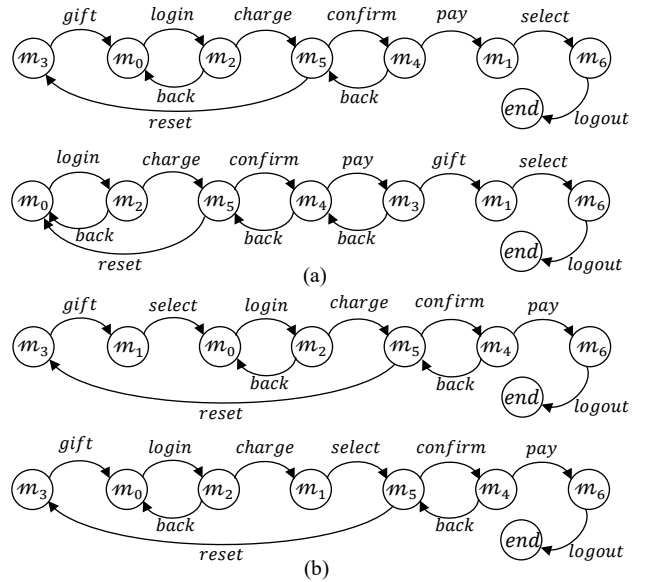

Fig. 12 Module-consistent revisions given by OACAL (a: two specifications selected from 46 eligible revisions, b: results after applying degradation)

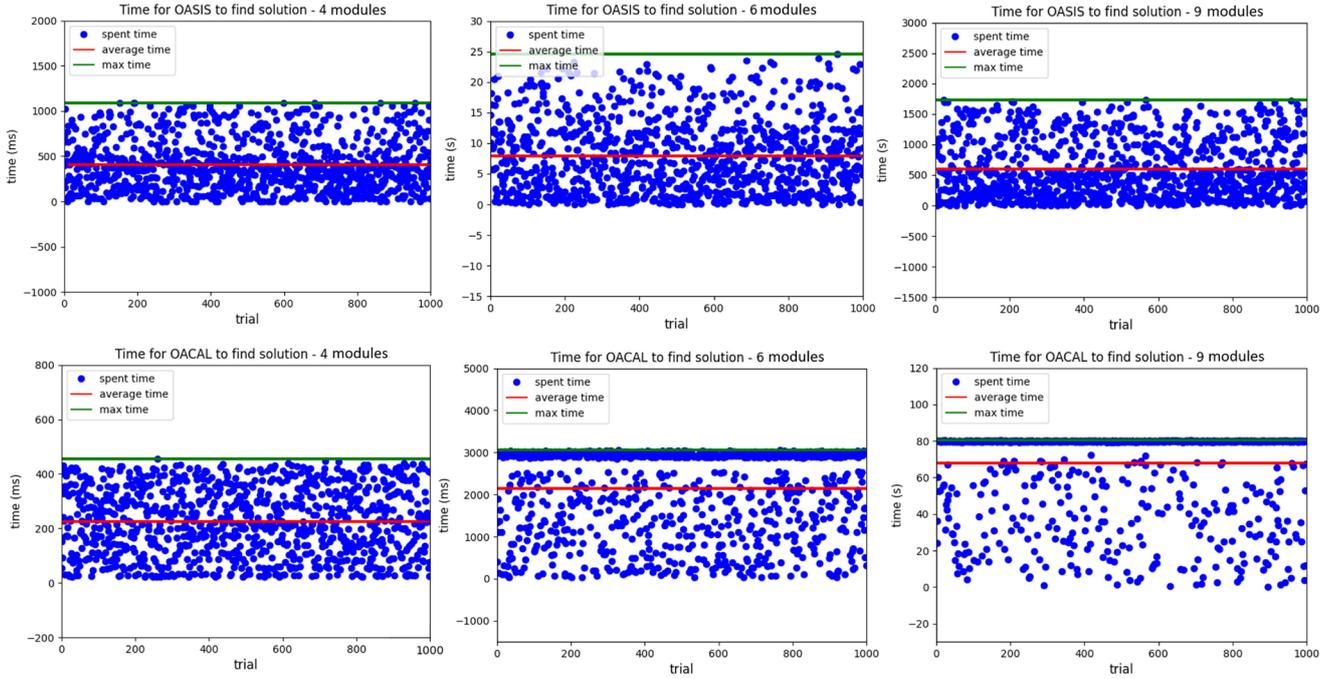
Fig. 13. Time consumption on finding eligible module-consistent revisions (up: OASIS, down: OACAL, from left to right: cases of 4/6/9 modules).

By comparing the quantity of the specification revisions generated by OASIS and OACAL, we demonstrated that more revisions could be found with larger coverage given by module-based recomposition, which answered RQ1. Moreover, we showed that more suitable revisions could be found by trading off insignificant requirements directly from the prediction results without the need to repeatedly run the algorithm, which responded to RQ2.

*C. Efficiency analysis*

For the general efficiency analysis of OASIS and OACAL, we followed the experimental setup in Fig. 9. We selected some representative results for the comparison of two approaches' searching speed in the same coverage, as shown in Fig. 13 where the horizontal axis is the number of trials of different requirements and the vertical axis is time in either millisecond or second. The ratio of the time consumption of OASIS to OACAL in average case and worst case is also plotted in Fig. 14. We have several notable findings illustrated as follows. (I) For OASIS, the data sparsity above the line of the average time is higher than the below one. (II) For OACAL, searching time are converged to the maximum time with the increment of number of modules. (III) OACAL has higher computational efficiency than OASIS; it performs much better with a larger number of modules and it is more advantageous for the worst case than the average case.

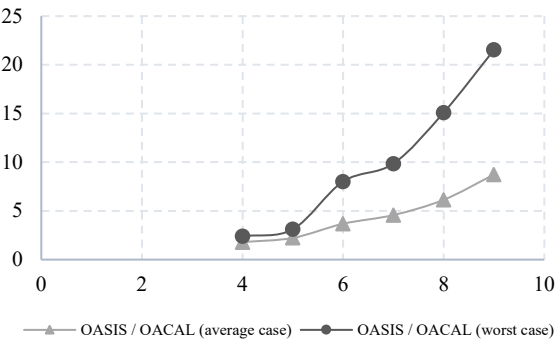
Fig. 14. Time consumption ratio of OASIS to OACAL.

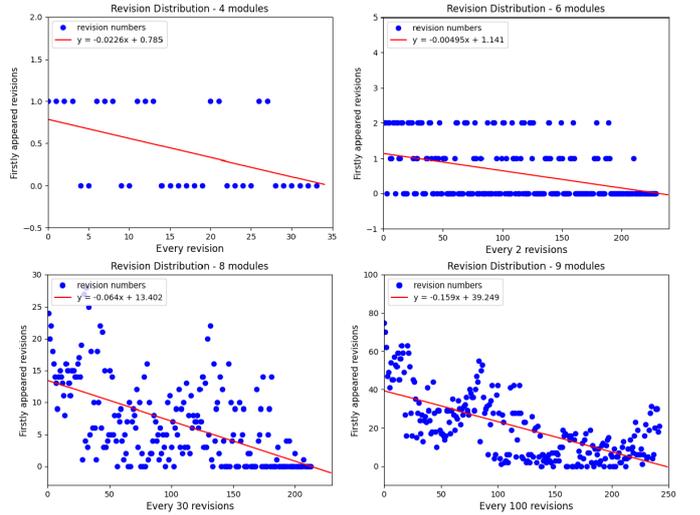
Fig. 15. The distribution of non-redundant module-consistent revisions in OASIS (horizontal axis: revisions in the coverage scaled by a certain number e.g., the distribution for 9 modules is scaled by 1/100 and 24309 revisions are checked in total, vertical axis: the number of revisions that first appear, plot: the number of the first appeared revisions in one scale, red line: linear regression result of the plots).

To investigate the reason for (I), we meticulously checked how non-redundant module-consistent revisions distribute in OASIS. Fig. 15 presents some representative results. The linear regression results indicate that more non-redundant revisions are covered earlier. Therefore, an eligible revision takes more time to be found with the increment of iterations, which explained the higher data sparsity above the average time in Fig. 13(a). To explain (II) and (III), we give a mathematical proof as follows. Equation (1) and (2) are the general formulas of the searching time for OASIS and OACAL respectively

$$t_i^{OASIS}(N, R_i) = pos_r(R_i) \cdot cov_r(N) \cdot t_{DCS}(N, R_i) \quad (1)$$

$$t_i^{OACAL}(N, R_i) = \begin{cases} pos_{nr}(R_i) \cdot cov_{nr}(N) \cdot t_{MC}(N, R_i) & , pos_{nr}(R_i) \leq \tau \\ \tau cov_{nr}(N) \cdot t_{MC}(N, R_i) + t_{ML}(N, R_i) & , pos_{nr}(R_i) > \tau \end{cases} \quad (2)$$

where $t_i^{OASIS}$ and $t_i^{OACAL}$ are the searching time of two approaches for the trial $i$. $R_i$ is the randomly generated requirements in the trial $i$ and $N$ is the number of modules. $pos_r$ and $pos_{nr}$ are the positions of revisions (dependent on $R_i$) in the coverage $cov_r$ that with revision redundancy and coverage $cov_{nr}$ that without redundancy (dependent on $N$). $t_{DCS}, t_{MC}, t_{ML}$ are the time consumption for each searching by controller synthesis, model checking and machine learning accordingly, which are dependent on $N$ and $R_i$. $\tau$ is the ratio of revisions split as training data for machine learning and is in the range of 0 to 1. When we implement OACAL, if an eligible revision is found in the model checking process ($pos_{nr}(R_i) \leq \tau$), it would return the result and not go further to the machine learning process. The value of $\tau$ needed for training an accurate model decreases with the increment of modules. Hence, when there are more modules in the specification, fewer eligible revisions can be found during the step of model checking, which explains (II).

In this paper, GBT is used as the classification algorithm, and its time complexity defined by XGBoost [38] (used in Turi Create) is $O(Kd\|\mathbf{x}_N\|_0 \log n)$ where $K$ is the number of trees (same as training epochs), $d$ is the maximum depth of trees (a constant), $\|\mathbf{x}_N\|_0$ is the $l_0$ norm [39] of each input $\mathbf{x}_N$, $n$ is the size of training data which could be obtained by multiplying split ratio $\tau$ and the size of the entire dataset. In this evaluation, $\|\mathbf{x}_N\|_0$ is equal to the $N$, since there is no zero in each input. We suppose that eligible revisions are uniformly distributed in the coverage so that the time consumption ratio of OASIS to OACAL would be given by (3) and (4) for average case and worst case respectively

$$\frac{t_{avg}^{OASIS}}{t_{avg}^{OACAL}} = \frac{\frac{1}{T}\sum_i^T(pos_r(R_i) \cdot cov_r(N)) \cdot t_{DCS}}{(\tau - \frac{1}{2}\tau^2) \cdot cov_{nr}(N) \cdot t_{MC} + (1-\tau)CKd\|\mathbf{x}_N\|_0 \log(\tau \cdot cov_{nr}(N))} \quad (3)$$

$$\frac{t_{max}^{OASIS}}{t_{max}^{OACAL}} = \frac{\max_{i \in (0,T)}(pos_r(R_i) \cdot cov_r(N)) \cdot t_{DCS}}{\tau \cdot cov_{nr}(N) \cdot t_{MC} + CKd\|\mathbf{x}_N\|_0 \log(\tau \cdot cov_{nr}(N))} \quad (4)$$

where $T$ is total trials and $C$ is a constant ratio of actual computational time to time complexity. We could interpret both equations as follows: the numerator is the theoretical average/maximum time for OASIS to find the revision where both $(1/T)\sum_i^T(pos_r(R_i) \cdot cov_r(N))$ and $\max(pos_r(R_i) \cdot cov_r(N))$ are the theoretical iterations presented in Fig. 4; the first term in the denominator is the model checking time where the non-redundant coverage $cov_{nr}(N)$ could be checked in Table I because this analysis is carried out in OASIS's coverage; and the second term in the denominator is the machine learning time. To examine $C$ on our device, we set a variety of complexity and checked the corresponding actual time, which resulted in $C \approx 1.12 \times 10^{-2}$. Then, we could verify the theoretical values for the time consumption ratios. We have $\tau = 0.3$, $K = 100$ and $d = 6$ by our setting. For example, when $N = 7$, the calculation results for (4) and (5) are $(471 t_{DCS})/(109.4 t_{MC} + 230.76)$ and $(1693 t_{DCS})/(128.7 t_{MC} + 329.65)$, respectively. $t_{DCS}$ and $t_{MC}$ would be the same since both OASIS and OACAL employ LTSA for controller synthesis and model checking. With $t_{DCS} = t_{MC} \approx 50$ milliseconds, we were able to acquire final values of 4.13 and 12.51, which are similar to the actual results in Fig. 14. As the module increases, the numerator of (4) would be boosted considerably faster than that of (3), as seen in Fig. 4. Also, as $\tau$ becomes smaller with more modules, the values of the denominators (i.e., OACAL's searching time) in (3) and (4) would be closer and always grow slower than the numerators. As a result, both (3) and (4) increase as the number of modules increases, with (4) growing faster than (3), which could explain (III).

Therefore, we answered RQ3 by demonstrating that OACAL outperforms OASIS in terms of the speed to find revised specifications with functional consistency, thanks to the removal of redundant revisions and the high efficiency of machine learning.

### D. Discussion

*Coverage of revisions.* By the case study, we successfully proved that OACAL could generate more module-consistent revisions than OASIS with the complete coverage given by the module-based recomposition. However, larger coverage for such type of revisions is not always better. Fig. 16 depicts a Venn diagram of the revision coverage by which four types of revisions are generalized. ① denotes the revisions covered by OASIS but not by OACAL, which do not maintain module consistency but remain the absolute partial order (i.e., composed by two refinements). ② as the intersection of OASIS and OACAL, includes the revisions that sustain both module-consistency and absolute partial order (the coverage we set for efficiency analysis). ③ contains the revisions that are module-consistent but do not absolutely maintain the partial order (e.g., revisions in Fig. 8 and Fig. 12). Lastly, ④ denotes the revisions that are out of the searching scope of both approaches and care neither of these two properties.

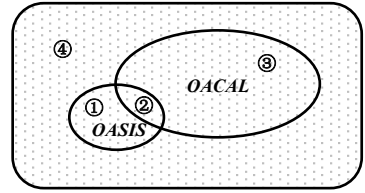

Fig. 16. Coverage of specification revisions

Thus, the decision between OACAL and OASIS is based on developer's preference for module-consistency and absolute partial order of the software specification. Moreover, OACAL would be a more efficient choice if both properties are considered significant, as proven by the analysis.

*Trade-offs.* Since we used machine learning to speed up the checking process, there is a trade-off between accuracy and efficiency in our method. Although the evaluation results demonstrate that the trained classification model has a promising accuracy, the accuracy cannot be guaranteed when machine learning is deployed in practice without any validation data. As a result, when compared to utilizing pure model checking, adopting machine learning implies losing the assurance of 100% accuracy. To address this issue, we could simply apply model checking to the prediction results to ensure their correctness. Another trade-off in our approach is between module consistency and revision flexibility. The OACAL approach focuses on the case that "$M_o$ does not refine $M_r$ and $\mathcal{A}_{M_r} = \mathcal{A}_{M_o}$", which does not allow any actions to be removed, added or replaced on the module. Thus, some revisions in ① and ④ of Fig. 16 are not covered, which might be the optimal choices. As a future work, we may introduce the degradation on module consistency to cover more potential eligible revisions.

## VI. Conclusion

We have proposed an efficient approach for automatically revising the software specification to make it more resilient to weakened user obligations that may lead to security threats. To achieve this goal, OACAL recomposes the behavior model of the specification without compromising functional consistency. We began by specifying several limitations of the most relevant work – OASIS, and then introduced our approach by explaining how each component could address those limitations. Later, we carried out a case study and a general efficiency analysis to demonstrate that OACAL outperforms OASIS in both coverage and searching speed of eligible module-consistent revisions. However, the decision between the two approaches is dependent on the developer's expectations for the system, as they focus different properties of the revision.

Our main future work is to validate OACAL by putting it through its paces in more realistic and complicated settings, as well as to introduce more flexible strategies (e.g., consistency degradation) for specification revision. In order to make further improvements, we will also evaluate the negative implications of implementing our approach in various real-world scenarios.

## References

[1] S. Sanath Kumar (2019). A study on customer satisfaction towards online ticket booking for airlines. *Journal of Information and Computational Science, 9*(8), 802-808.

[2] Mingyue Zhang and Shuxian Li. 2020. Study on the tourism behavior preference of Full Nest I families: Demand for Online Booking Platform. In *Proceedings of the 2020 2nd International Conference on Big Data and Artificial Intelligence (ISBDAI '20)*. Association for Computing Machinery, New York, NY, USA, 191–197. DOI: https://doi.org/10.1145/3436286.3436322

[3] Irwin, K., Yu, T., & Winsborough, W. H. (2006, October). On the modeling and analysis of obligations. In *Proceedings of the 13th ACM conference on Computer and communications security* (pp. 134-143).

[4] Andrada F. (2020, November 9). *From pranks to APTs: How remote access Trojans became a major security threat*. CSO. https://www.csoonline.com/article/3588156

[5] Tun, T. T., Bennaceur, A., & Nuseibeh, B. (2020, August). *OASIS: Weakening User Obligations for Security-critical Systems. In 2020 IEEE 28th International Requirements Engineering Conference (RE)* (pp. 113-124). IEEE.

[6] Jeff M., Jeff K., Robert C., Sebastain U., Howard F. (2006 June). *LTSA - Labelled Transition System Analyser*. https://www.doc.ic.ac.uk/ltsa/

[7] Turi. (2018). *Graphlab Create. Fast, Scalable Machine Learning Modeling in Python*. https://turi.com/.

[8] Keller, R. M. (1976). Formal verification of parallel programs. *Communications of the ACM, 19*(7), 371-384.

[9] Pnueli, A. (1977, October). The temporal logic of programs. In *18th Annual Symposium on Foundations of Computer Science (sfcs 1977)* (pp. 46-57). IEEE.

[10] Rozier, K. Y. (2011). Linear temporal logic symbolic model checking. *Computer Science Review, 5*(2), 163-203.

[11] Baier, C., & Katoen, J. P. (2008). *Principles of model checking*. MIT press.

[12] Giannakopoulou, D., & Magee, J. (2003, September). Fluent model checking for event-based systems. In *Proceedings of the 9th European software engineering conference held jointly with 11th ACM SIGSOFT international symposium on Foundations of software engineering* (pp. 257-266).

[13] Yee, K. P. (2004). Aligning security and usability. *IEEE Security & Privacy, 2*(5), 48-55.

[14] Zurko, M. E., & Simon, R. T. (1996, September). User-centered security. In *Proceedings of the 1996 workshop on New security paradigms* (pp. 27-33).

[15] Brachmann, E., Dittmann, G., & Schubert, K. D. (2012, September). Simplified authentication and authorization for restful services in trusted environments. In *European Conference on Service-Oriented and Cloud Computing* (pp. 244-258). Springer, Berlin, Heidelberg.

[16] Florencio, D., & Herley, C. (2007, May). A large-scale study of web password habits. In *Proceedings of the 16th international conference on World Wide Web* (pp. 657-666).

[17] Shay, R., Komanduri, S., Durity, A. L., Huh, P., Mazurek, M. L., Segreti, S. M., ... & Cranor, L. F. (2016). Designing password policies for strength and usability. *ACM Transactions on Information and System Security (TISSEC), 18*(4), 1-34.

[18] Möller, S., Ben-Asher, N., Engelbrecht, K. P., Englert, R., & Meyer, J. (2011). Modeling the behavior of users who are confronted with security mechanisms. *Computers & Security, 30*(4), 242-256.

[19] Katsabas, D., Furnell, S., & Dowland, P. (2005, July). Using human computer interaction principles to promote usable security. In *Proceedings of the Fifth International Network Conference (INC 2005)*, Samos, Greece (pp. 235-242).

[20] Naqvi, B., & Seffah, A. (2018, May). A methodology for aligning usability and security in systems and services. In *2018 3rd International Conference on Information Systems Engineering (ICISE)* (pp. 61-66). IEEE.

[21] Sasse, M. A., Brostoff, S., & Weirich, D. (2001). Transforming the 'weakest link'—a human/computer interaction approach to usable and effective security. *BT technology journal, 19*(3), 122-131.

[22] Bishop, M. (2005) Psychological Acceptability Revisited. Cranor, L, F., Garfinkel, S. *Security and usability: designing secure systems that people can use* (pp. 1 - 12).. " O'Reilly Media, Inc.".

[23] Martinelli, F., & Matteucci, I. (2007). An approach for the specification, verification and synthesis of secure systems. *Electronic Notes in Theoretical Computer Science*, 168, 29-43.

[24] Braberman, V., D'Ippolito, N., Piterman, N., Sykes, D., & Ucriitel, S. (2013, May). Controller synthesis: From modelling to enactment. In *2013 35th International Conference on Software Engineering (ICSE)* (pp. 1347-1350). IEEE.

[25] Munir, A., & Koushanfar, F. (2018). Design and analysis of secure and dependable automotive CPS: A steer-by-wire case study. *IEEE Transactions on Dependable and Secure Computing, 17*(4), 813-827.

[26] Sander, O., Klimm, A., & Becker, J. (2013). Hardware support for authentication in cyber physical systems.

[27] Magee, J., & Kramer, J. (1999). *State models and java programs. wiley Hoboken, 10*, 332036.

[28] Cattani, G. L., & Sewell, P. (2000). *Models for name-passing processes: Interleaving and causal*. Cambridge [Cambridgeshire]: University of Cambridge, Computer Laboratory.

[29] D'Ippolito, N., Braberman, V., Kramer, J., Magee, J., Sykes, D., & Uchitel, S. (2014, May). Hope for the best, prepare for the worst: multi-tier control for adaptive systems. In *Proceedings of the 36th International Conference on Software Engineering* (pp. 688-699).

[30] Cai, C. H., Sun, J., & Dobbie, G. (2019). Automatic B-model repair using model checking and machine learning. *Automated Software Engineering, 26*(3), 653-704.

[31] Hendriks, M. (2019). *Can machine learning automatically choose your best model checking strategy?* (Bachelor's thesis, University of Twente).

[32] Amrani, M., Lúcio, L., & Bibal, A. (2018). ML+ FV= ♥? A Survey on the Application of Machine Learning to Formal Verification. *arXiv preprint arXiv:1806.03600*.

[33] Zhu, W., Wu, H., & Deng, M. (2019). LTL model checking based on binary classification of machine learning. *IEEE Access, 7*, 135703-135719.

[34] *Support Streamers by Cheering with Bits!* (n.d.). Twitch. https://www.twitch.tv/bits.

[35] *LIVE gifting*. (2021). TikTok. https://www.tiktok.com/creators/creator-portal/en-us/getting-paid-to-create/live-gifting/.

[36] *GIFT – Send gifts to creators to make videos better*. (n.d.). Niconico. https://nicoad.nicovideo.jp/gift.

[37] *What are B-Coin, Coin or Membership used for?* (2018, October). Bilibili. https://www.bilibili.com/read/cv1305402.

[38] Chen, T., & Guestrin, C. (2016, August). Xgboost: A scalable tree boosting system. In *Proceedings of the 22nd acm sigkdd international conference on knowledge discovery and data mining* (pp. 785-794).

[39] Wang, Y., Zhang, H., Chen, H., Boning, D., & Hsieh, C. J. (2020). On $l_p$-norm Robustness of Ensemble Stumps and Trees. *arXiv preprint arXiv:2008.08755*.